\documentclass[english,aps,preprint]{revtex4-1}
\usepackage[T1]{fontenc}
\usepackage[latin9]{inputenc}
\usepackage{xcolor}
\usepackage{pdfcolmk}
\usepackage{amsmath}
\usepackage{amssymb}
\usepackage{graphicx}
\PassOptionsToPackage{normalem}{ulem}
\usepackage{ulem}

\makeatletter

\providecolor{lyxadded}{rgb}{0,0,1}
\providecolor{lyxdeleted}{rgb}{1,0,0}

\@ifundefined{showcaptionsetup}{}{%
 \PassOptionsToPackage{caption=false}{subfig}}
\usepackage{subfig}
\makeatother

\usepackage{babel}
\begin{document}

\title{Evaluation of Spectral Zeta-Functions with the Renormalization Group}

\author{Stefan Boettcher and Shanshan Li}

\affiliation{Department of Physics, Emory University, Atlanta, GA 30322; USA }
\begin{abstract}
We evaluate spectral zeta-functions of certain network Laplacians
that can be treated exactly with the renormalization group. As specific
examples we consider a class of Hanoi networks and those hierarchical
networks obtained by the Migdal-Kadanoff bond moving scheme from regular
lattices. As possible applications of these results we mention quantum
search algorithms as well as synchronization, which we discuss in
more detail.
\end{abstract}
\maketitle

\section{Introduction\label{sec:Introduction}}

Spectral zeta-functions have numerous applications in many areas of
mathematics \citep{Voros92} and the sciences \citep{Ramon97,dunne2012KernelFractals}.
Especially notable examples of recent use in physics concern the synchronization
dynamics on complex networks \citep{Barahona02,Korniss03} or quantum
search algorithms \citep{Childs04}. In both cases, the spectral zeta-function
pertains to properties of a lattice Laplacian. For the former case,
the zeta function becomes a stand-in to approximate the smallest nontrivial
Laplacian eigenvalue, in the latter, it allows us to relate the spectral
dimension of the network to the computational complexity of quantum
search.

Here, we study these spectral zeta-functions using exact renormalization
group methods. To this end, we employ classes of hierarchical networks
that exhibit geometric as well as small-world properties. Hierarchies
in various forms \citep{Simon62,Southern77,Hoffmann88,SWPRL,Agliari15}
have a number of useful functions while describing a range of behaviors,
from lattice-like to mean field. The Hanoi networks \citep{SWPRL,Boettcher09c,Boettcher10c}
mix a geometric structure, a one-dimensional loop, with small-world
bonds in a tractable, recursive manner. They have been used recently
to demonstrate explosive percolation in hierarchical networks \citep{Singh14,Boettcher11d},
as well as to design new, synthetic phase transitions for various
spin models \citep{Singh14b,BoBr12}. In turn, the Migdal-Kadanoff
renormalization group (MKRG) \citep{Migdal76,Kadanoff76} has already
a venerable history, with countless results to successfully describe
the phase diagrams of finite-dimensional systems in statistical \citep{Plischke94,Pathria},
condensed matter \citep{Berker79,F+H}, and particle physics \citep{itzykson1989statistical}.
MKRG provides an effective way to explore the phase diagram of systems
on $d$-dimensional lattices. It is particularly useful as a complement
to mean-field theory for understanding the properties of lattice models
in low dimensions. We find highly nontrivial results for the scaling
properties of their Laplacian determinants, as an extension of our
previous work \citep{BoLi15}, and in the case of MKRG we can analytically
continue results to entire families of lattice models. Elsewhere \citep{LiBo16},
we show, how this work can be used, for instance, to predict the efficiency
of quantum search as a function of the spectral dimension. Here, we
focus specifically on applications to synchronization.

This paper is organized as follows: in Sec.\ \ref{sec:Graph-Structure},
we describe the structure and properties of hierarchical networks
in which we study spectral zeta functions; in Sec.\ \ref{sec:Spectral-Determinants},
we introduce the spectral zeta functions applied in various scenarios
and its evaluation via a heuristic argument; in Sec.\ \ref{sec:RG-Calculation},
we outline the renormalization group procedure on the Hanoi networks
for the evaluations of Laplacian determinant and spectral zeta functions;
in Sec.\ \ref{sec:RG-for-MKRG}, we derive the RG recursions and
spectral zeta function in hierarchical networks from MKRG; in Sec.\ \ref{sec:Conclusion},
we conclude by applying the spectral zeta function to describe synchronization.
In the Appendix we also apply RG to the power method to determine
the largest eigenvalue of the Laplacian for all the networks we consider,
as needed for synchronization. Many other details of our investigations
are also explained in the Appendix. 

\section{Network Structure\label{sec:Graph-Structure}}

\subsection{Hanoi Networks\label{subsec:Hanoi-Networks}}

\begin{figure}
\includegraphics[bb=100bp 100bp 400bp 700bp,clip,angle=270,scale=0.5]{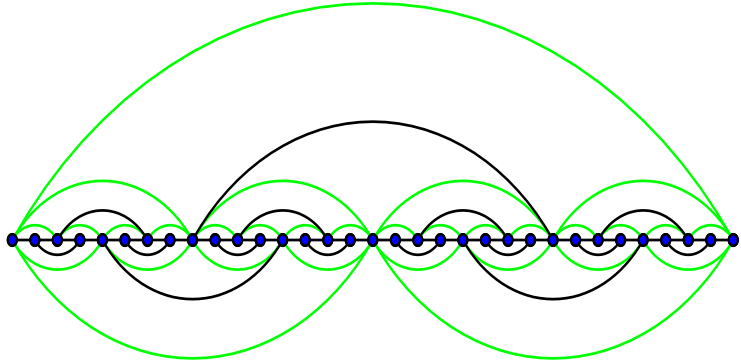}

\caption{\label{fig:5hanoi}Depiction of the Hanoi networks HN3 (black bonds
only) and HN5 (black and green-shaded bonds). }
\end{figure}

The hanoi networks \citep{SWPRL,Boettcher09c,Boettcher10c,Singh14b}
possess a simple geometric backbone, a one-dimensional line of $N=2^{k}$
sites. Each site is at least connected to its nearest neighbor left
and right on the backbone. To generate the small-world hierarchy in
these networks, consider parameterizing any number $n<N$ (except
for zero) \emph{uniquely} in terms of two other integers $(i,j)$,
$i\geq1$ and $1\leq j\leq2^{k-i}$, via
\begin{eqnarray}
n & = & 2^{i-1}\left(2j-1\right).\label{eq:numbering}
\end{eqnarray}
Here, $i$ denotes the level in the hierarchy whereas $j$ labels
consecutive sites within each hierarchy. To generate the network HN3,
we connect each site $n=2^{i-1}(4j-3)$ also with a long-distance
neighbor $n'=2^{i-1}(4j-1)$ for $1\leq j\leq2^{k-i-1}$, as shown
in Fig. \ref{fig:5hanoi}. While it is of a fixed, finite degree,
we can extend HN3 in the following manner to obtain a new network
of average degree 5, called HN5. In addition to the bonds in HN3,
in HN5 we also connect all even sites to both of its nearest neighboring
sites \emph{within} the same level of the hierarchy $i\geq1$ in Eq.\ (\ref{eq:numbering}).
The resulting network remains planar but now sites have a hierarchy-dependent
degree with an exponential degree distribution, also demonstrated
in Fig.\ \ref{fig:5hanoi}. Previously\citep{SWPRL}, it was found
that the average chemical path between sites on HN3 scales as $d^{HN3}\sim\sqrt{N}$,
reminiscent of a square-lattice consisting of $N$ lattice sites.
In HN5, it is easy to show recursively that this distance grows as
$d^{HN5}\sim\log_{2}N$ \citep{Boettcher10c}.

\subsection{Migdal-Kadanoff renormalization group\label{sec:Diamond-Fractals}}

\begin{figure*}
\includegraphics[bb=0bp 200bp 558bp 805bp,clip,width=0.24\textwidth]{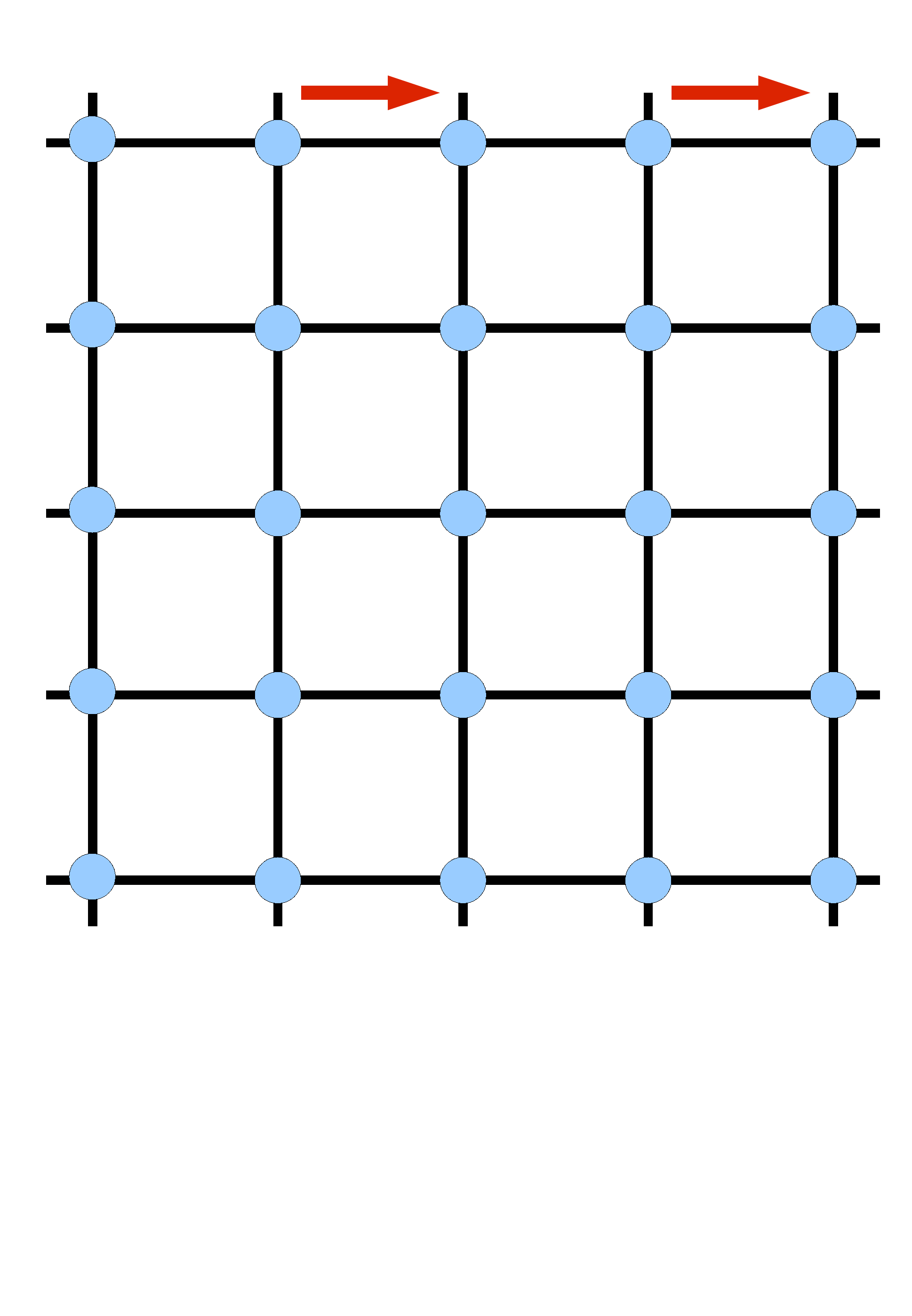}\includegraphics[bb=0bp 200bp 558bp 805bp,clip,width=0.24\textwidth]{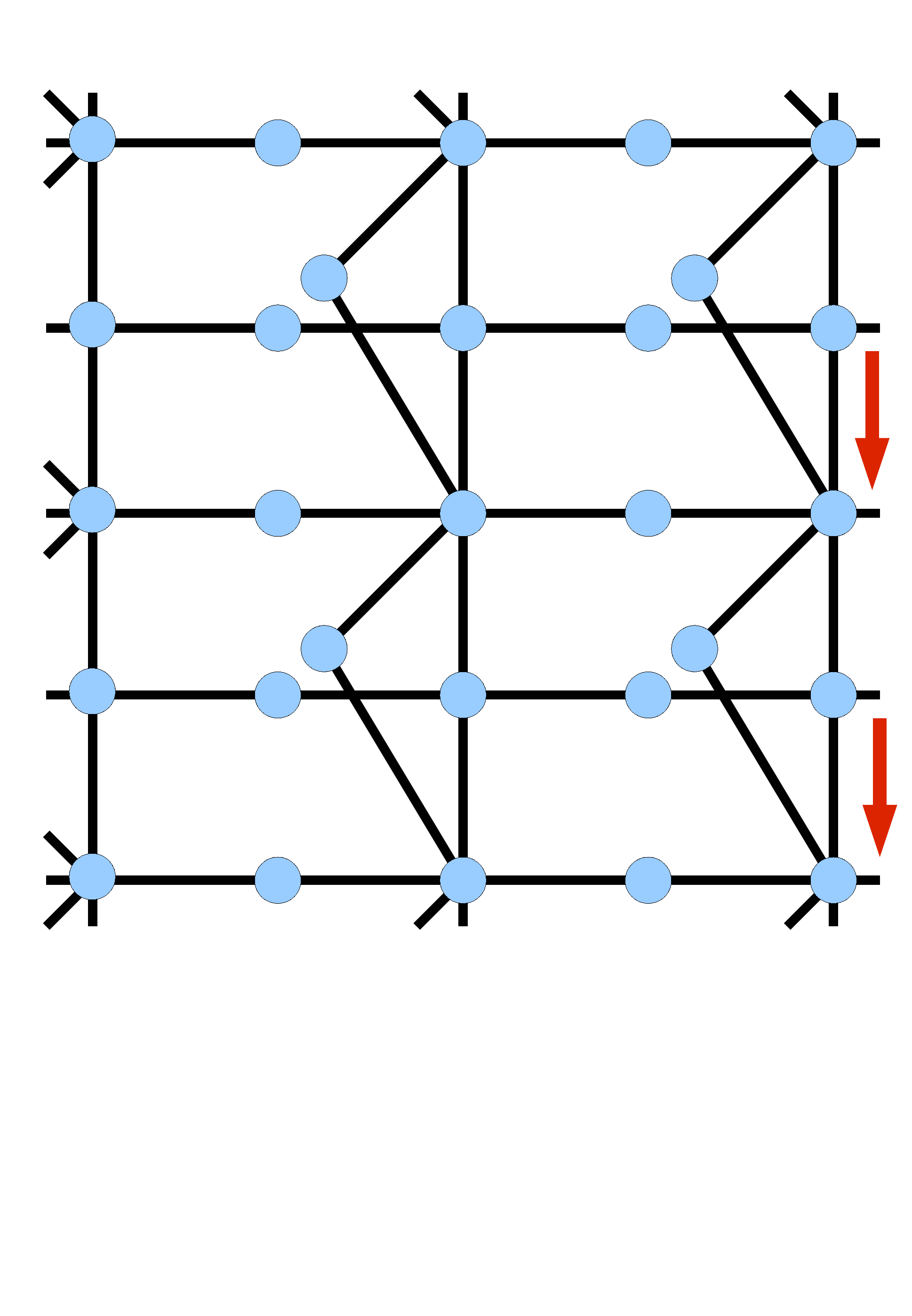}\includegraphics[bb=0bp 200bp 558bp 805bp,clip,width=0.24\textwidth]{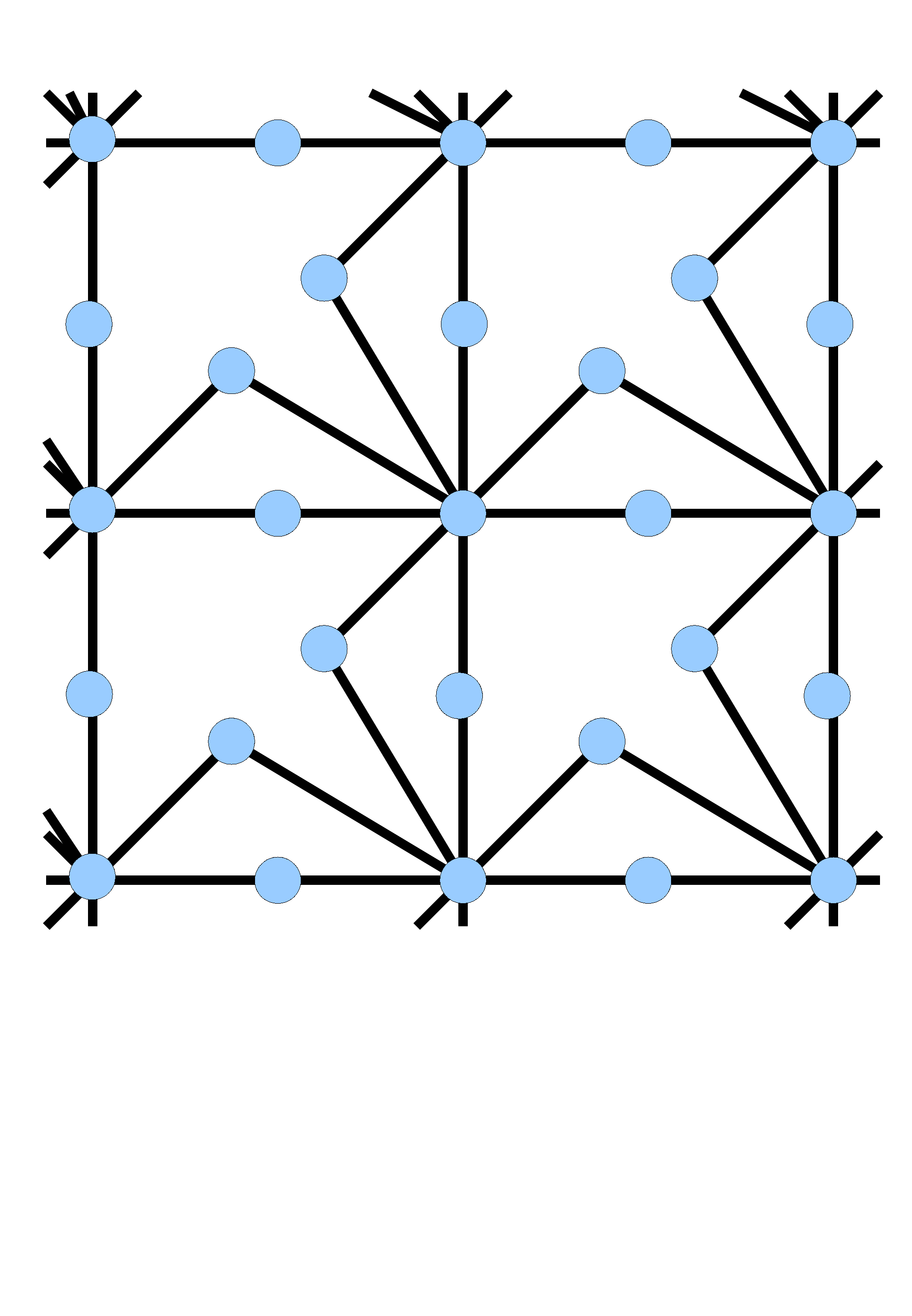}\includegraphics[bb=263bp 217bp 530bp 516bp,clip,width=0.22\textwidth]{MKlattice1}\caption{\label{fig: MKlattice} Bond-moving scheme in the Migdal\textendash Kadanoff
renormalization group, here for a square lattice ($d=2$) with $l=2$,
i.~e. $b=2$ in Eq.~(\ref{eq:bl}). Starting from the lattice with
unit bonds (a), bonds in intervening hyper-planes are projected onto
every $l^{{\rm th}}$ plane in one direction while bonds connect to
the $l^{{\rm th}}$ plane only at every $l^{{\rm th}}$ vertex (b),
which is then repeated in subsequent directions (c), to re-obtain
a similar hyper-cubic lattice, now of bond-length $l$ (d). The renormalized
bonds in this case consist of $b=l^{d-1}=2$ branches, each of a series
of $l=2$ bonds; the general RG-step for $l=2$ and arbitrary branches
$b$ is depicted in Fig. \ref{fig:MKRG}}
\end{figure*}

\begin{figure}
\includegraphics[bb=50bp 220bp 270bp 600bp,clip,width=0.3\textwidth]{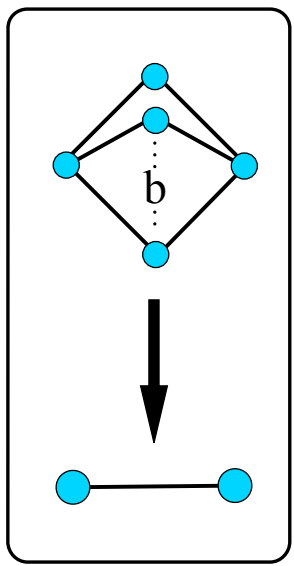}\caption{\label{fig:MKRG}Hierarchical RG that results from the bond-moving
scheme of any $d$-dimensional lattice shown in Fig. \ref{fig: MKlattice}.
A collection of $b$ strings of $l$ bonds at generation $\mu$ each
($l=2$ here) gets renormalized into a single new bond at generation
$\mu+1$. }
\end{figure}

The Migdal-Kadanoff renormalization group (MKRG)~\citep{Migdal76,Kadanoff76,Berker79}
is a bond-moving scheme that approximates $d$-dimensional lattices.
It often provides excellent approximations for $d=2$ and 3 \citep{Southern77},
and it becomes trivially exact in $d=1.$ The networks resulting from
MKRG have a simple recursive, yet geometric, structure and have been
widely studied in statistical physics~\citep{F+H,Plischke94,Pathria}.
Starting from generation $\mu$ with a single bond, at each subsequent
generation $\mu+1,$ all bonds are replaced with a new sub-graph.
This structure of the sub-graph arises from the bond-moving scheme
in $d$ dimensions~\citep{Migdal76,Kadanoff76}, as depicted in Fig.~\ref{fig: MKlattice}:
In a hyper-cubic lattice of unit bond length, at first all $l-1$
intervening hyper-planes of bonds, transverse to a chosen direction,
are projected into every $l^{{\rm th}}$ hyper-plane, followed by
the same step for $l-1$ hyper-planes being projected onto the $l^{{\rm th}}$
plane in the next direction, and so on. In the end, as shown in Fig.
\ref{fig:MKRG}, one obtains a renormalized hyper-cubic lattice (of
bond length $l)$ in generation $\mu+1$ with a renormalized bond
of generation $\mu+1$ consisting of a sub-graph of 
\begin{equation}
b=l^{d-1}\label{eq:bl}
\end{equation}
 parallel branches, each having of a series of $l$ bonds of generation
$\mu$. In turn, we can rewrite Eq.~(\ref{eq:bl}) as 
\begin{equation}
d=1+\log_{l}b,\label{eq:deq}
\end{equation}
anticipating analytic continuation in $l$ and $b$ to obtain results
for arbitrary, \emph{real} dimensions $d$. In the following, we consider
a general series of Migdal-Kadanoff networks by varying $b$ while
fixing $l=2$. 

\section{Spectral Zeta-Functions of Laplacians\label{sec:Spectral-Determinants}}

The Laplacian matrix is given by 
\begin{eqnarray}
\left[{\bf L}\right]_{i,j} & = & d_{i}\delta_{i,j}-A_{i,j},\label{eq:Laplacian}
\end{eqnarray}
where $d_{i}$ specifies the degree of the $i$-th vertex and $A_{i,j}$
is the adjacency matrix of the network. Since the links in the networks
are undirected, ${\bf A}$ and ${\bf L}$ are symmetric. By design,
all row or column sums in ${\bf L}$ vanish, i.e., $\sum_{i}\left[{\bf L}\right]_{i,j}=\sum_{j}\left[{\bf L}\right]_{i,j}=0$.
The fundamental property of the Laplacian matrix is its spectrum of
eigenvalues, the solutions $\lambda_{i}$ of the secular equation
\begin{eqnarray}
\det\left[{\bf L}-\lambda{\bf 1}\right] & = & 0.\label{eq:secular}
\end{eqnarray}
With an RG approach \citep{BoLi15}, the effort of determining the
spectrum reduces \emph{exponentially} from solving $2^{k}\times2^{k}$
determinants to $k$ iterations in a few RG recursion equations for
any desired quantity.

We motivate our studies into the spectrum of the Laplacian matrix
and their spectral zeta-functions through the intimate connection
between various dynamic properties of transport phenomena and the
geometry expressed via the Laplacian. In particular, it has been shown
that the ratio between lowest and highest (nontrivial) eigenvalue
provides a measure for the synchronization ability of coupled identical
oscillators located on the nodes of the network\citep{Barahona02}.
Another synchronization problem emerges in the context of parallel
discrete-event simulations (PDES) \citep{Korniss03,Kozma04}, where
nodes must frequently ``synchronize'' with their neighbors (on a
given network) to ensure causality in the underlying simulated dynamics.
The local synchronizations, however, can introduce correlations in
the resulting synchronization landscape, leading to strongly nonuniform
progress at the individual processing nodes. The above is a prototypical
example for synchronization in many systems such as causally constrained
queuing networks, supply-chain networks based on electronic transactions
\citep{Nagurney05}, etc.

Consider an arbitrary network in which the nodes interact through
the links. The nodes are assumed to be task processing units, such
as computers or manufacturing devices. Each node has completed an
amount of task $h_{i}$ and these together at all nodes constitute
the task-completion (synchronization) landscape $\{h_{i}(t)\}_{i=1}^{N}$.
Here $t$ is the discrete number of parallel steps executed by all
nodes, which is proportional to the real time, and $N$ is the number
of nodes. In this particular model the nodes whose local field variables
are incremented by an exponentially distributed random amount at a
given step are those whose completed task amount is not greater than
the tasks at their neighbors. Thus, denoting the neighborhood of the
node $i$ by $S_{i}$, if $h_{i}(t)\leq\min_{j\in S_{i}}\{h_{j}(t)\}$,
the node $i$ completes some additional exponentially distributed
random amount of task; otherwise, it idles. In its simplest form the
evolution equation for the amount of task completed at the node $i$
can be written as 
\begin{equation}
h_{i}(t+1)=h_{i}(t)+\eta_{i}(t)\prod_{j\in S_{i}}\Theta\left(h_{j}(t)-h_{i}(t)\right),\label{eq_motion}
\end{equation}
where $h_{i}(t)$ is the local field variable (amount of task completed)
at node $i$ at time $t$; $\eta_{i}(t)$ are \emph{iid} random variables
with unit mean, delta-correlated in space and time (the new task amount);
and $\Theta(...)$ is the Heaviside step function. Despite its simplicity,
this rule preserves unaltered the asynchronous dynamics of the underlying
system. The larger the disparity in task completion is, the more memory
has to be stored in the advanced units, which is costly in the context
of limited resources. A measure of that cost, then, is the amount
of de-synchronization, which is provided by the average ``surface-roughness''
\begin{eqnarray}
\left\langle w^{2}\right\rangle  & = & \frac{1}{N}\sum_{i=2}^{N}\frac{1}{\lambda_{i}},\label{eq:w2}
\end{eqnarray}
where $\lambda_{i}$ are the rank-ordered eigenvalues of the Laplacian
matrix ${\bf L}$ of the network, leaving out the trivial, lowest
eigenvalue $\lambda_{1}=0$. It is very difficult to analytically
calculate each eigenvalue individually to evaluate the sum defining
$\left\langle w^{2}\right\rangle $ in Eq.\ (\ref{eq:w2}). 

As similar problem is encountered in the evaluation of the efficiency
of quantum search on a network \citep{Childs04}. However, in this
case, we need to access even higher moments of the eigenvalues. So,
it becomes useful to define an entire function generating such moments:
\begin{equation}
I_{j}=\frac{1}{N}\,\sum_{i=2}^{N}\left(\frac{1}{\lambda_{i}}\right)^{j},\label{SZF}
\end{equation}
the spectral zeta-function \citep{Voros92,akkermans2009physical,dunne2012KernelFractals}.
Note, for instance, that in the evaluation of partition functions
in field theory the $I_{j}$ often feature in the continuation to
non-integer moments, in particular, the limit $j\to0$ \citep{Ramon97}.
In almost all cases, with the exception of regular lattices where
Fourier transforms can be applied, and some fractals \citep{Rammal84},
it is impossible to find each eigenvalue in the sum of Eq.\ (\ref{SZF}).
However, the sum defined in Eq.\ (\ref{SZF}) for $I_{j}$ can be
expressed as the $j^{{\rm th}}$ derivative of the determinant ${\bf \mathbf{\mathcal{L}}}+\epsilon\mathbb{I}$
in the limit $\epsilon\to0$: 

\begin{eqnarray}
I_{j} & = & \frac{1}{N}\,\left.\frac{\left(-1\right)^{j-1}}{\left(j-1\right)!}\left(\frac{\partial}{\partial\epsilon}\right)^{j}\,\sum_{i=2}^{N}\ln\left(\lambda_{i}+\epsilon\right)\right|_{\epsilon\to0},\nonumber \\
 & = & \frac{1}{N}\,\left.\frac{\left(-1\right)^{j-1}}{\left(j-1\right)!}\left(\frac{\partial}{\partial\epsilon}\right)^{j}\,\ln\left[\frac{1}{\epsilon}\prod_{i=1}^{N}\left(\lambda_{i}+\epsilon\right)\right]\right|_{\epsilon\to0},\nonumber \\
 & = & \frac{1}{N}\,\left.\frac{\left(-1\right)^{j-1}}{\left(j-1\right)!}\left(\frac{\partial}{\partial\epsilon}\right)^{j}\,\ln\left[\frac{1}{\epsilon}\det\left({\bf L}+\epsilon{\bf 1}\right)\right]\right|_{\epsilon\to0},\label{eq:Izetadet}
\end{eqnarray}
where we have used the fact that $\lambda_{1}=0$. This has the advantage
that we do not need to know each individual Laplacian eigenvalue $\lambda_{i}$,
as has been previously assumed in the context of quantum search and
many other applications\citep{Agliari2011,Agliari16}. Note, for instance,
that Eq. (\ref{eq:w2}) now reduces to $\left\langle w^{2}\right\rangle =I_{1}$.
In the following, we can take advantage of the RG-techniques developed
for Laplacian determinants in Ref. \citep{BoLi15} to derive the scaling
of $I_{j}$.

\subsection{A Heuristic Argument\label{subsec:Heuristic-Arguments}}

\begin{figure}
\includegraphics[bb=80bp 0bp 600bp 720bp,clip,angle=270,width=0.49\columnwidth]{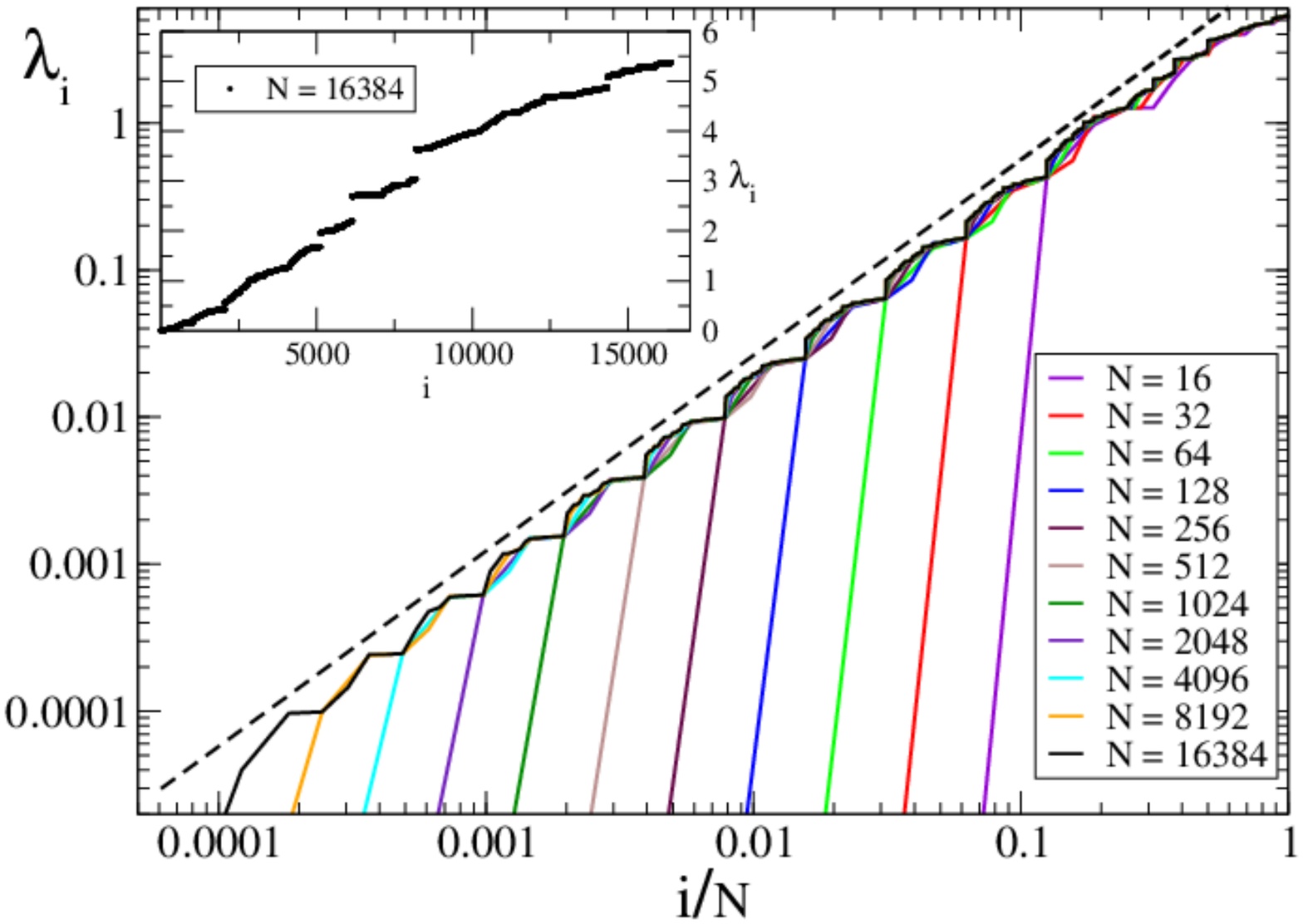}\hfill{}\includegraphics[bb=80bp 0bp 600bp 720bp,clip,angle=270,width=0.49\columnwidth]{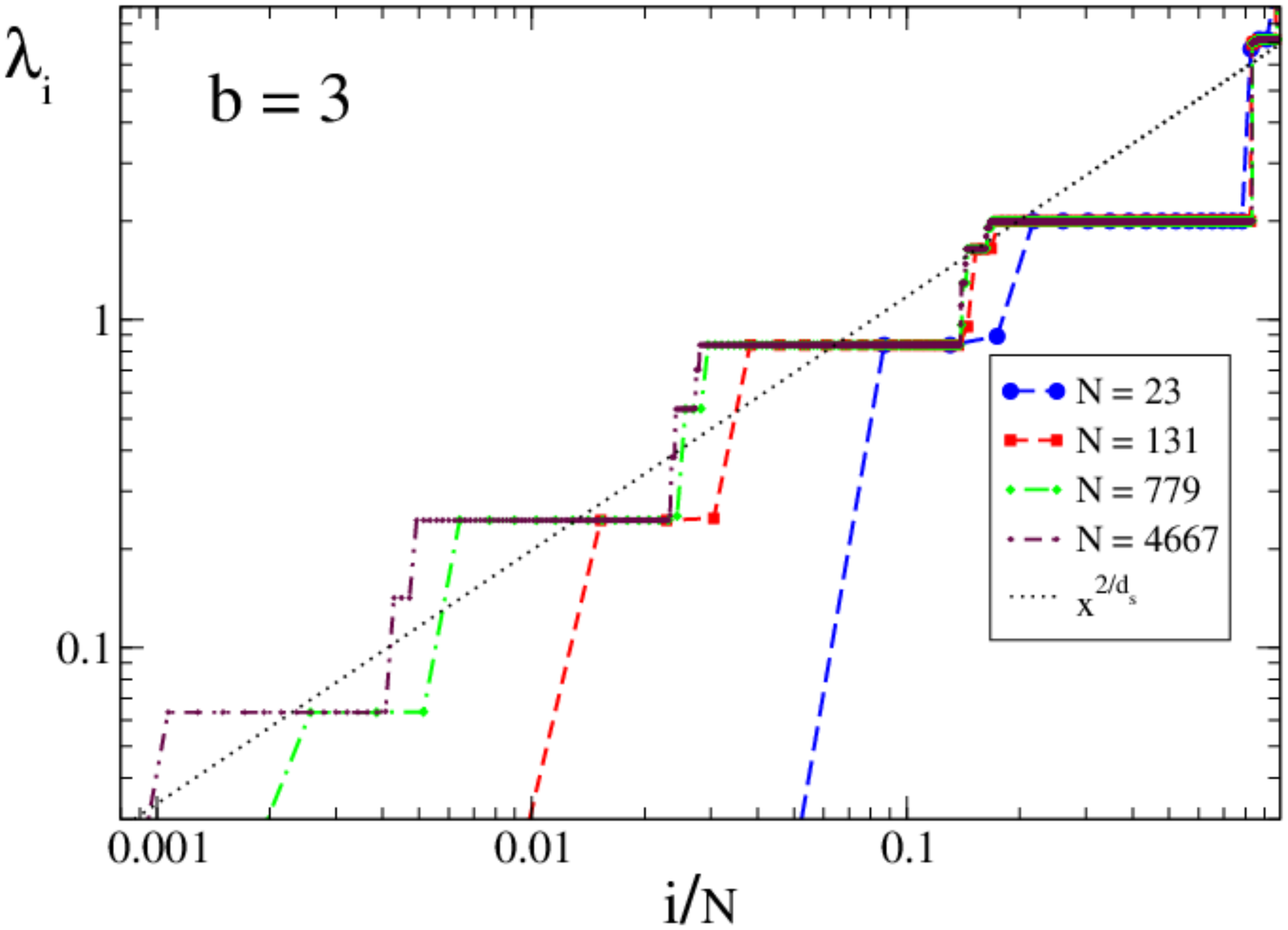}\caption{\label{fig:HN3spectra}Plot of the Laplacian eigenvalues $\lambda_{i}$
ordered by rank $i$ as a function of $\frac{i}{N}$ for various system
sizes $N$ of the Hanoi network HN3 (left) and MKRG at $b=3$ (right).
The dashed line follows a power law with exponent $2/d_{s}=d_{w}/d_{f}\approx1.31$
\citep{SWN} for HN3, and for MKRG with $d_{s}=1+\log_{2}3$. The
inset for HN3 shows the same data for only $N=2^{16}$ on a linear
scale, showing that the scaling concerns only the part of the spectrum
where $\frac{i}{N}$ is small. While the assumptions underlying Eq.\ (\ref{eq:lambdax})
seem well-justified for HN3, the high degree of degeneracy in MKRG
spectra makes the heuristic argument in Sec.\ \ref{subsec:Heuristic-Arguments}
more dubious, and progressively worse for larger $b$.}
\end{figure}

If we assume \citep{Rammal84} that the rank-ordered eigenvalues $0=\lambda_{1}<\lambda_{i}\leq\lambda_{i_{c}}$
for all $2\leq i\leq i_{c}$ up to some $2\leq i_{c}\leq N$ for large
$N$ follow a power-law form,
\begin{equation}
\lambda_{i}\sim\left(\frac{i}{N}\right)^{\frac{2}{d_{s}}},\label{eq:lambdax}
\end{equation}
and $\lambda_{i}\sim const$ for $i>i_{c}$. This is shown, for instance,
in Fig.\ \ref{fig:HN3spectra} to be applicable for the fractal network
HN3, for which $2/d_{s}=d_{w}/d_{f}=2-\log_{2}\phi\approx1.31$ with
$\phi=\left(\sqrt{5}+1\right)/2$ \citep{SWN}, but it is at best
vaguely satisfied for MKRG even in the best-case scenario, $b=3$,
because of an ever higher degree of degeneracy in the spectrum. Note
that under these assumptions, it is in fact easy to evaluate the spectral
zeta-function in Eq.\ (\ref{SZF}) directly by taking the Riemann
limit, $\frac{i}{N}\to\theta$ with $d\theta=\frac{1}{N}$, such that
\begin{eqnarray}
\frac{1}{N}\,\sum_{i=2}^{N}\left(\frac{1}{\lambda_{i}}\right)^{j} & \sim & \int_{\frac{1}{N}}^{\theta_{c}}d\theta\,\theta^{-\frac{2j}{d_{s}}}+const,\nonumber \\
 & \sim & N^{\frac{2j}{d_{s}}-1}+const.\label{eq:Nds}
\end{eqnarray}
This result would hold for any $\theta_{c}=\frac{i_{c}}{N}\sim N^{-\nu}$
with $0\leq\nu<1$, such that the $N$-dependent scaling is dominated
by the lower limit of the integral.

\section{RG for the Spectral Determinant of Hanoi Networks \label{sec:RG-Calculation}}

The determinant of ${\cal L}(\epsilon)={\bf L}+\epsilon{\bf 1}$ in
Eq.\ (\ref{eq:Izetadet}) for fractal lattices can be evaluated asymptotically
in a recursive renormalization scheme. We have already described the
procedure in great detail in Ref. \citep{BoLi15}. Here we only outline
the procedure to be able to focus on the novel aspects need for our
calculation here. In general, we employ the well-known formal identity~\citep{Ramon97},
\begin{eqnarray}
\frac{1}{\sqrt{\det{\cal L}}} & = & \idotsint_{-\infty}^{\infty}\left(\prod_{i=1}^{N}\frac{dx_{i}}{\sqrt{\pi}}\right)\exp\left\{ -\sum_{n=1}^{N}\sum_{m=1}^{N}x_{n}{\cal L}_{n,m}x_{m}\right\} .\label{eq:gaussint}
\end{eqnarray}
For the RG, we employ a hierarchical scheme by which at each step
$\mu$ a fraction $1/b$ of all remaining variables get integrated
out while leaving the integral in Eq.\ (\ref{eq:gaussint}) invariant,
but now with $N^{\prime}\leq N/b$ variables. Formally, say, in case
of $b=2$ we integrate out every odd-indexed variable in a network
at step $\mu$, we separate $\prod_{i=1}^{N}dx_{i}=\prod_{i=1}^{N/2}dx_{2i}\prod_{j=1}^{N/2}dx_{2j+1}$
and integrate to receive
\begin{equation}
\frac{1}{\sqrt{\det{\cal L}}}=C^{\prime}\idotsint_{-\infty}^{\infty}\left(\prod_{i=1}^{\frac{N}{2}}\frac{dx_{2i}}{\sqrt{\pi}}\right)\exp\left\{ -\sum_{n=1}^{\frac{N}{2}}\sum_{m=1}^{\frac{N}{2}}x_{2n}{\cal L}_{n,m}^{\prime}x_{2m}\right\} ,
\end{equation}
where the reduced Laplacian ${\cal L}^{\prime}$ is now a $\frac{N}{2}\times\frac{N}{2}$
matrix that is formally \emph{identical} with ${\cal L}$ and $C^{\prime}$
is an overall scale-factor. That is, if ${\cal L}={\cal L}\left(q,p,\ldots\right)$
depends on some parameters, then ${\cal L}^{\prime}={\cal L}^{\prime}\left(q^{\prime},p^{\prime},\ldots\right)$
depends on those parameters in the same functional form, thereby revealing
the RG-recursion relations, $q^{\prime}=q^{\prime}\left(q,p,\ldots\right)$,
$p^{\prime}=p^{\prime}\left(q,p,\ldots\right)$, etc, and $C^{\prime}=C^{\prime}\left(q,p,\ldots\right)$,
that encapsulate all information of the original Laplacian. After
a sufficient number of such RG-steps, a reduced Laplacian of merely
a few variables remains that can be solved by elementary means. This
property, of course, is very special and can be iterated in exact
form only for certain types of fractal networks.

In Ref. \citep{BoLi15}, we have shown, for example, that for the
Hanoi networks HN3 and HN5 we find the RG recursions: 
\begin{eqnarray}
q_{\mu+1} & = & q_{\mu}+2l_{\mu}-2\frac{p_{\mu}^{2}}{q_{\mu}-1},\nonumber \\
p_{\mu+1} & = & l_{\mu}+\frac{p_{\mu}^{2}}{q_{\mu}-1},\label{eq:RG-HN3_redux}\\
l_{\mu+1} & = & l_{0}+\frac{p_{\mu}^{2}}{q_{\mu}^{2}-1},\nonumber 
\end{eqnarray}
 and 
\begin{eqnarray}
C_{k}^{(\mu)} & = & \prod_{i=0}^{\mu-1}\left[q_{i}^{2}-1\right]^{-2^{k-3-i}},\label{eq:Ikproduct}
\end{eqnarray}
such that the determinant of the Laplacian after $k-2$ RG-steps becomes:

\begin{eqnarray}
\det\left[{\bf {\bf L}_{k}^{(3,5)}}+\epsilon{\bf 1}\right] & \sim & \epsilon\left[C_{k}^{(k-2)}\right]^{-2}.\label{eq:secularHN3}
\end{eqnarray}
Note that the termination condition for the final RG-step merely contribute
a factor of $\sim\epsilon$ that is needed to cancel the $1/\epsilon$
in Eq.\ (\ref{eq:Izetadet}) due to the $\lambda_{1}=0$-eigenvalue.
The asymptotic behavior of the determinant itself arises entirely
from $C_{k}^{(k-2)}$. Only the initial conditions on the RG-recursions
distinguish between HN3 and HN5. These are:
\begin{eqnarray}
C_{k}^{(0)} & = & 1,\nonumber \\
q_{0} & = & 3+\epsilon,\label{eq:IC_HN5}\\
p_{0} & = & 1,\nonumber \\
l_{0} & = & \begin{cases}
0, & {\rm for\,\,HN3}\\
1, & {\rm for\,\,HN5}
\end{cases}.\nonumber 
\end{eqnarray}
As shown in Ref. \citep{BoLi15} (and easily verified by insertion),
the parameters $\left\{ q_{k-2},p_{k-2},l_{k-2}\right\} $ in Eq.(\ref{eq:RG-HN3_redux})
in HN3 approach fixed points as 

\begin{align}
q_{\mu} & \sim1+\left(\frac{2}{\phi}\right)^{-\mu}\left(Q_{0}+\epsilon\left(\frac{4}{\phi}\right)^{\mu}Q_{1}+\epsilon^{2}\left(\frac{4}{\phi}\right)^{2\mu}Q_{2}\ldots\right),\nonumber \\
p_{\mu} & \sim\left(\frac{2}{\phi}\right)^{-\mu}\left(P_{0}+\epsilon\left(\frac{4}{\phi}\right)^{\mu}P_{1}+\epsilon^{2}\left(\frac{4}{\phi}\right)^{2\mu}P_{2}\ldots\right),\label{eq:Ansatz-HN3}\\
l_{\mu} & \sim\left(\frac{2}{\phi}\right)^{-\mu}\left(L_{0}+\epsilon\left(\frac{4}{\phi}\right)^{\mu}L_{1}+\epsilon^{2}\left(\frac{4}{\phi}\right)^{2\mu}L_{2}\ldots\right),\nonumber 
\end{align}
where $\phi=\left(\sqrt{5}+1\right)/2$. In turn, the set of parameters
in Eq.(\ref{eq:RG-HN3_redux}) for HN5 approach the fixed points as 

\begin{align}
q_{\mu} & \sim\frac{5+\sqrt{41}}{2}+\left(\epsilon2^{\mu}Q_{1}+\epsilon^{2}2^{2\mu}Q_{2}\ldots\right),\nonumber \\
p_{\mu} & \sim\frac{3+\sqrt{41}}{4}+\left(\epsilon2^{\mu}P_{1}+\epsilon^{2}2^{2\mu}P_{2}\ldots\right),\label{eq:Ansatz-HN5}\\
l_{\mu} & \sim\frac{3+\sqrt{41}}{8}+\left(\epsilon2^{\mu}L_{1}+\epsilon^{2}2^{2\mu}L_{2}\right).\nonumber 
\end{align}

When Eq.\ (\ref{eq:secularHN3}) is evaluated for HN3, the overall
factor $C_{k}^{\left(k-2\right)}$ is approximated using $q_{\mu}$
in Eq. (\ref{eq:Ansatz-HN3}),

\begin{eqnarray}
C_{k}^{(\mu)} & \sim & \alpha^{N}\prod_{i=0}^{\mu-1}\left[\left(1+\left(\frac{2}{\phi}\right)^{-i}\left(Q_{0}+\epsilon\left(\frac{4}{\phi}\right)^{i}Q_{1}+\epsilon^{2}\left(\frac{4}{\phi}\right)^{2i}Q_{2}\ldots\right)\right)^{2}-1\right]^{-2^{k-3-i}},\label{eq:Ikproduct-approx}
\end{eqnarray}
 in which the parameter $\alpha$ is determined to any accuracy by
simple iteration of the recursions in Eq. (\ref{eq:RG-HN3_redux}),
which was executed in Ref. \citep{BoLi15}. For HN3, $\alpha=2.0189990298\ldots$;
for HN5, $\alpha=2.7548806715\ldots$. However, the existence of the
factor $\alpha^{N}$ is irrelevant for the scaling of $I_{j}$ since
it has no contribution to the derivative of the \emph{logarithm} of
the determinant with respect to $\epsilon$. Applying Eq. (\ref{eq:Izetadet}),
the zeta functions for HN3 eventually read as

\begin{eqnarray}
I_{j} & \sim & \left[\left(\frac{2}{\phi}\right)^{\log_{2}N}\right]^{j},\nonumber \\
 & \sim & \left[N^{1-\log_{2}\phi}\right]^{j}.\label{eq:zetaFunctionHN3}
\end{eqnarray}
Similarly, inserting $q_{i}$ in Eq. (\ref{eq:Ansatz-HN5}) into Eqs.
(\ref{eq:Ikproduct}-\ref{eq:secularHN3}) provides for HN5: 

\begin{eqnarray}
I_{j} & \sim & \begin{cases}
\begin{array}{c}
\log_{2}N\\
N^{2j/2-1}
\end{array} & \begin{array}{c}
j=1,\\
j\geq2.
\end{array}\end{cases}\label{eq:zetaFunctionHN5}
\end{eqnarray}

For HN3 at $j=1$ in Eq. (\ref{eq:zetaFunctionHN3}), it is easy to
identify the exponent as $1-\log_{2}\phi=d_{w}-d_{f}$, in which $d_{w}=2-\log_{2}\phi$
is the random walk dimension obtained for HN3 in Ref. \citep{SWN}
in the metric where the $1d$-backbone defines distances such that
$d_{f}=1$. The results for the spectral zeta-function is consistent
with that found generally by Ref. \citep{Giacometti95},
\begin{equation}
\left\langle w^{2}\right\rangle \sim L^{d_{w}-d_{f}},
\end{equation}
 for the surface roughness defined in Eq.\ (\ref{eq:w2}) for which
we have shown in Sec. \ref{sec:Spectral-Determinants} that $I_{1}=\left\langle w^{2}\right\rangle $.
Using $d_{w}=2d_{f}/d_{s}$ \citep{Alexander82} and the definition
$N=L^{d_{f}}$ then leads to
\begin{equation}
I_{j}\sim N^{\frac{2j}{d_{s}}-1},\label{eq:IjN}
\end{equation}
for $d_{s}<2j$, uniquely described in terms of the spectral dimension.
For HN5, the spectral dimension is $d_{s}=2$, leading to the logarithmic
scaling in Eq.\ (\ref{eq:zetaFunctionHN5}) for $j=1$ where $d_{s}\geq2j$.
When $j\geq2$, Eq.\ (\ref{eq:IjN}) also applies to HN5. 

\begin{figure}
\includegraphics[bb=0bp 0bp 330bp 612bp,clip,width=0.3\textwidth]{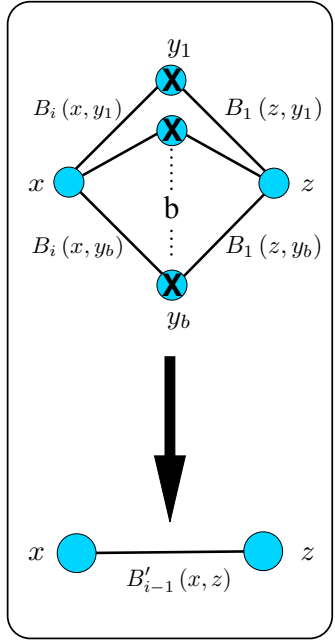}\caption{\label{fig:MKRGzeta}Graph-let for the MKRG for the spectral zeta-function,
as adapted from the generic structure shown in Fig. \ref{fig:MKRG}.
In that graph-let, the $b$ inner vertices $y_{1},\ldots,y_{b}$ belong
to the currently lowest level ($i=0$) of the hierarchy that will
be integrated out (${\bf \times}$-mark) in the next RG-step. One
of the two outer vertices, $z$, must be exactly one level higher
($i=1$, here shown right). The other outer vertex, $x$, must be
of some unspecified but higher level ($i>1$, left). After the RG-step,
symbolized by the arrow, the renormalized link $B_{i-1}^{\prime}$
is bound to have a vertex $z$ with $i=0$ on one end and some vertex
$x$ with $i^{\prime}=i-1>0$ on the other. A set of $2b$ of these
links then become the input of the \textendash{} identical \textendash{}
next RG-step.}
\end{figure}

\section{RG for the Spectral Determinant of Migdal-Kadanoff\label{sec:RG-for-MKRG}}

Since we have not considered MKRG before in this way, we derive its
RG-recursions here in more detail. To this end, we can reconstruct
the integral in Eq.\ (\ref{eq:gaussint}) piece-by-piece by defining
a simple algebra. As suggested by Fig.\ \ref{fig: MKlattice}(c),
in each RG-step the lattice consists of a collection of graph-lets
of the type shown in Fig.\ \ref{fig:MKRG}, which we have adapted
for the following calculation in Fig. \ref{fig:MKRGzeta}. In that
graph-let, the $b$ inner vertices belong to the currently lowest
level ($i=0$) of the hierarchy that will be integrated out in the
next RG-step. One of the two outer vertices is exactly one level higher
($i=1$) as it would be integrated at the next step. The other outer
vertex must be of some unspecified but higher level ($i>1$). We can
now define a helpful function pertaining to each bond, each of which
is bound to have a vertex with $i=0$ on one end and some vertex with
$i>0$ on the other. Its part of the integrand in Eq.\ (\ref{eq:gaussint})
has the form
\begin{equation}
B_{i}\left(x,y\right)=C_{i}\exp\left\{ -\frac{q_{i}}{2}x^{2}-\frac{q_{0}}{2}y^{2}+2pxy\right\} ,\label{eq:DefB}
\end{equation}
such that the RG-step depicted in Fig.\ \ref{fig: MKlattice}(d)
amounts to
\begin{eqnarray}
B_{i-1}^{\prime}\left(x,z\right) & = & \idotsint_{-\infty}^{\infty}\prod_{j=1}^{b}\frac{dy_{j}}{\sqrt{\pi}}\,B_{i}\left(x,y_{j}\right)B_{1}\left(z,y_{j}\right),\nonumber \\
 & = & C_{i}^{b}C_{1}^{b}\exp\left\{ -\frac{b}{2}\left(q_{i}x^{2}+q_{1}z^{2}\right)\right\} \idotsint_{-\infty}^{\infty}\prod_{j=1}^{b}\frac{dy_{j}}{\sqrt{\pi}}\,\exp\left\{ -q_{0}y_{j}^{2}+2p\left(x+z\right)y_{j}\right\} ,\nonumber \\
 & = & C_{i}^{b}C_{1}^{b}q_{0}^{-\frac{b}{2}}\exp\left\{ -\frac{b}{2}\left(q_{i}-\frac{2p^{2}}{q_{0}}\right)x^{2}-\frac{b}{2}\left(q_{1}-\frac{2p^{2}}{q_{0}}\right)z^{2}+2b\frac{p^{2}}{q_{0}}xz\right\} ,\nonumber \\
 & = & C_{i-1}^{\prime}\exp\left\{ -\frac{q_{i-1}^{\prime}}{2}x^{2}-\frac{q_{0}^{\prime}}{2}z^{2}+2p^{\prime}xz\right\} ,\label{eq:B_RGstep}
\end{eqnarray}
where unprimed parameters are $\mu$-times previously renormalized
while primes indicate newly $\mu+1$-times renormalized parameters.
From the last two lines, we can read off the RG-recursions at the
$\mu^{{\rm th}}$ step:
\begin{eqnarray}
C_{i-1}^{(\mu+1)} & = & \left(\frac{C_{1}^{(\mu)}C_{i}^{(\mu)}}{\sqrt{q_{0}^{(\mu)}}}\right)^{b},\nonumber \\
q_{i-1}^{(\mu+1)} & = & b\,\left(q_{i}^{(\mu)}-\frac{2\left(p^{(\mu)}\right)^{2}}{q_{0}^{(\mu)}}\right),\label{eq:MKRGmu}\\
p^{(\mu+1)} & = & b\,\frac{\left(p^{(\mu)}\right)^{2}}{q_{0}^{(\mu)}},\nonumber 
\end{eqnarray}
for $i>0$. Considering that initially, at $\mu=0$ in the unrenormalized
network, all vertex-weights defined in Eq.\ (\ref{eq:DefB}) are
the same, $q_{i}^{(0)}\equiv2$ for all $i$, the distinction between
levels $i$ in Eq.\ (\ref{eq:MKRGmu}) disappears. Note that a vertex
at level $i>0$ contributes to the Gaussian integral $2b^{i}$-fold
through respective factors $B_{i}$, and 2-fold for $i=0$ by appearing
in two such factors $B_{i^{\prime}}$, $i^{\prime}>0$. In this manner,
the lattice Laplacian at $\mu=0$ in Eq.\ (\ref{eq:gaussint}) receives
its proper weights on its diagonal. Equally, $C_{i}^{(0)}=1$ for
all $i>0$. Thus, defining $C_{\mu}=C_{i}^{(\mu)}$, $p_{\mu}=b^{-\mu}p^{(\mu)}$,
and $q_{\mu}=b^{-\mu}q_{i}^{(\mu)}$ for all $i\geq0$, we obtain:
\begin{eqnarray}
C_{\mu+1} & = & \left[\frac{C_{\mu}^{2}}{\sqrt{b^{\mu}q_{\mu}}}\right]^{b},\qquad\left(C_{0}=1\right),\nonumber \\
q_{\mu+1} & = & q_{\mu}-2\frac{p_{\mu}^{2}}{q_{\mu}},\qquad\left(q_{0}=2-\epsilon\right),\label{eq:MKRG_redux}\\
p_{\mu+1} & = & \frac{p_{\mu}^{2}}{q_{\mu}},\qquad\left(p_{0}=1\right).\nonumber 
\end{eqnarray}
Note that the recursions in Eq.\ (\ref{eq:MKRG_redux}) is not exactly
identical to Eq.\ (\ref{eq:MKRGmu}). With eigenvalue $\lambda=-\epsilon$,
the initial condition for $q_{i}^{\left(0\right)}$ Eq.\ (\ref{eq:MKRGmu})
is, in fact,

\begin{eqnarray}
q_{i}^{\left(0\right)} & = & 2+\epsilon/b^{i},\qquad0\leq i<k,\label{eq:MKRG-initial}\\
q_{k}^{\left(0\right)} & = & 2+2\epsilon/b^{k},\qquad i=k,\nonumber 
\end{eqnarray}
which does not allow to collapse the $i$-th hierarchy like in the
Hanoi networks. However, in the Taylor expansion in small $\epsilon$,
order-by-order such a collapse is allowed. The difference between
the $\left\{ q_{0}^{\left(\mu\right)},p^{\left(\mu\right)}\right\} $
from Eq.\ (\ref{eq:MKRG_redux}) and $\left\{ q_{\mu},p_{\mu}\right\} $
from Eq.\ (\ref{eq:MKRGmu}) is

\begin{eqnarray}
q_{0}^{\left(\mu\right)}-q_{\mu} & \sim & Q_{1}\epsilon+Q_{2}\epsilon^{2}+Q_{3}\epsilon^{3}+\ldots,\label{eq:Diff-q_0-q}\\
p^{\left(\mu\right)}-p_{\mu} & \sim & P_{1}\epsilon+P_{2}\epsilon^{2}+P_{3}\epsilon^{3}+\ldots
\end{eqnarray}
 in which coefficients are all constants dependent only on the parameter
$b$. After $k-1$ iterations, the network is renormalized to two
end nodes, the Laplacian determinant is 

\begin{eqnarray}
\det\left[{\bf {\bf L}_{k}^{MK}}+\epsilon{\bf 1}\right] & = & C_{k}^{-2}b^{2k}\,\det\left[\begin{array}{cc}
q_{k}/2 & -p_{k}\\
-p_{k} & q_{k}/2
\end{array}\right]\nonumber \\
 & = & C_{k}^{-2}b^{2k}\left(q_{k}^{2}/4-p_{k}^{2}\right),\label{eq:SecularMK}
\end{eqnarray}
where the $C_{k}^{-2}$ can be expressed in closed form,

\begin{eqnarray*}
C_{k}^{-2} & = & \left(b^{0}q_{0}\right)^{2^{k-1}b^{k}}\left(b^{1}q_{1}\right)^{2^{k-2}b^{k-1}}\left(b^{2}q_{2}\right)^{2^{k-3}b^{k-2}}\ldots\left(b^{k-1}q_{k-1}\right)^{2^{k-k}b}\\
 & = & \left(\prod_{\mu=0}^{k-1}b^{\left(2b\right)^{k-\mu}\thinspace\mu/2}\right)\left(\prod_{\mu=0}^{k-1}q_{\mu}^{b\thinspace\left(2b\right)^{k-1-\mu}}\right).
\end{eqnarray*}
The ansatz for fixed points of rescaled $\left\{ q_{\mu},p_{\mu}\right\} $
in Eq.\ (\ref{eq:MKRG_redux}) is 

\begin{align}
q_{\mu} & \sim2^{-\mu}\left(Q_{0}+\epsilon4^{\mu}Q_{1}+\epsilon^{2}4^{2\mu}Q_{2}\ldots\right),\nonumber \\
p_{\mu} & \sim2^{-\mu}\left(Q_{0}/2-\epsilon4{}^{\mu}P_{1}/4+\epsilon^{2}4^{2\mu}P_{2}+\ldots\right).\label{eq:D_b_2-fps}
\end{align}
The fixed point scaling of parameters $q_{\mu}$ and $p_{\mu}$ in
Eq.\ (\ref{eq:D_b_2-fps}) verifies the validity of approximations
in Eq.\ (\ref{eq:MKRG_redux}), since the differences between the
approximated and exact parameters in Eq.\ (\ref{eq:Diff-q_0-q})
will not affect the scaling of any quantity we consider in Eq. (\ref{eq:Izetadet}).
With respect to $\epsilon$, we can calculate the $j^{{\rm th}}$
derivative of determinant for any $b$. Note that the asymptotic expression
for $\left[C_{k}^{(k-1)}\right]^{-2}$ is approximated to 

\begin{eqnarray*}
C_{k}^{-2} & \sim\alpha^{N} & \prod_{\mu=0}^{k-1}\left[q_{\mu}\right]^{b\thinspace\left(2b\right)^{k-1-\mu}},
\end{eqnarray*}
 in which $\alpha$ is determined respectively as $1.0594630943\ldots$,
$1.0233738919\ldots$, $1.0124545480\ldots$, $1.0077313692\ldots$,
and $1.0052649262\ldots$ for $b=2,3,4,5,6$. As argued above, however,
any such $\epsilon$-independent factor remains irrelevant after the
differentiation in Eq. (\ref{eq:Izetadet}).

The zeta-functions for the Laplacian determinants with varying $b$
are eventually evaluated as 

\begin{eqnarray}
I_{j} & \sim & \begin{cases}
\begin{array}{c}
N^{2j/(1+\log_{2}b)-1}\\
\ln N\\
const
\end{array} & \begin{array}{c}
2j>(1+\log_{2}b)\\
2j=(1+\log_{2}b)\\
2j<(1+\log_{2}b)
\end{array}\end{cases}\label{eq:IjN-MKRG}
\end{eqnarray}
Considering that the spectral dimensions for MKRG with $l=2$\citep{akkermans2009physical}
are 

\begin{eqnarray}
d_{s} & =d= & 1+\log_{2}b,\label{eq:d_s-MKRG}
\end{eqnarray}
 the zeta-functions are again identified as 

\begin{eqnarray}
I_{j} & \sim & N^{2j/d_{s}-1},\qquad2j>d_{s},\label{eq:MKRG-ds_Ij}
\end{eqnarray}
as in Eq, (\ref{eq:IjN}).

\section{Conclusion\label{sec:Conclusion}}

\begin{figure*}
\includegraphics{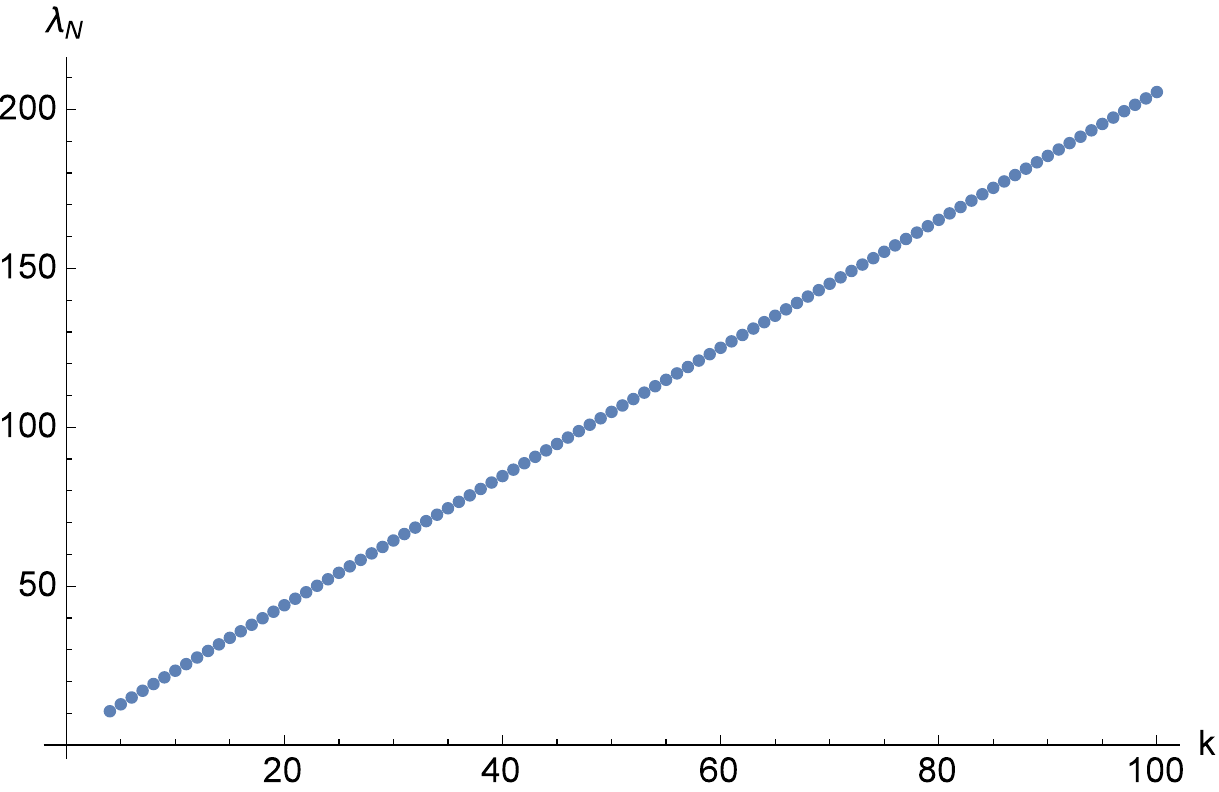}

\caption{\label{fig: maxEigenvalue-HN5} Plot of the largest eigenvalue $\lambda_{N}$
for HN5 with system size $k\left(=\log_{2}N\right)$, as obtained
by the power-method using the recursions in Eq. (\ref{eq:PMrecur-Hn5})
. A linear fit provides a scaling of $\lambda_{N}\sim2.023588046\ldots k+3.460393100\ldots$.}
\end{figure*}

\begin{figure}
\subfloat[b=2]{\includegraphics[width=0.45\textwidth]{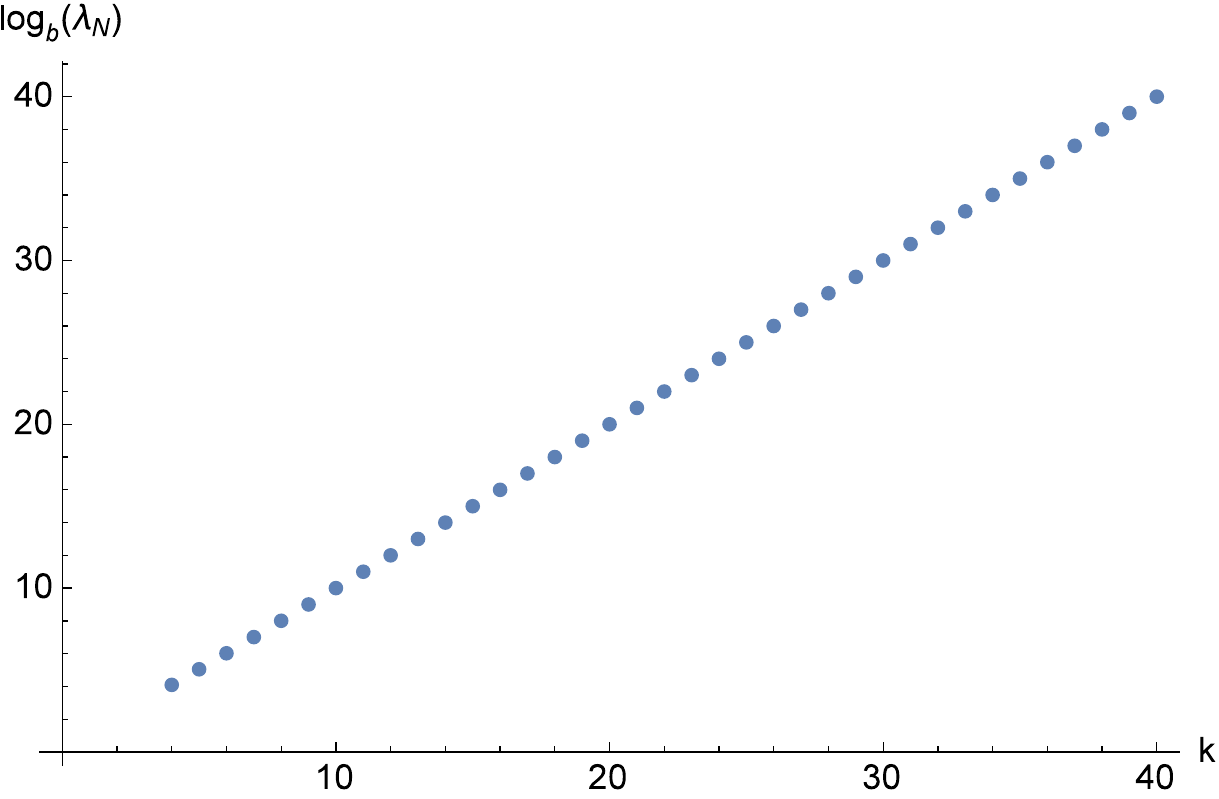}

}\qquad{}\subfloat[b=3]{\includegraphics[width=0.45\textwidth]{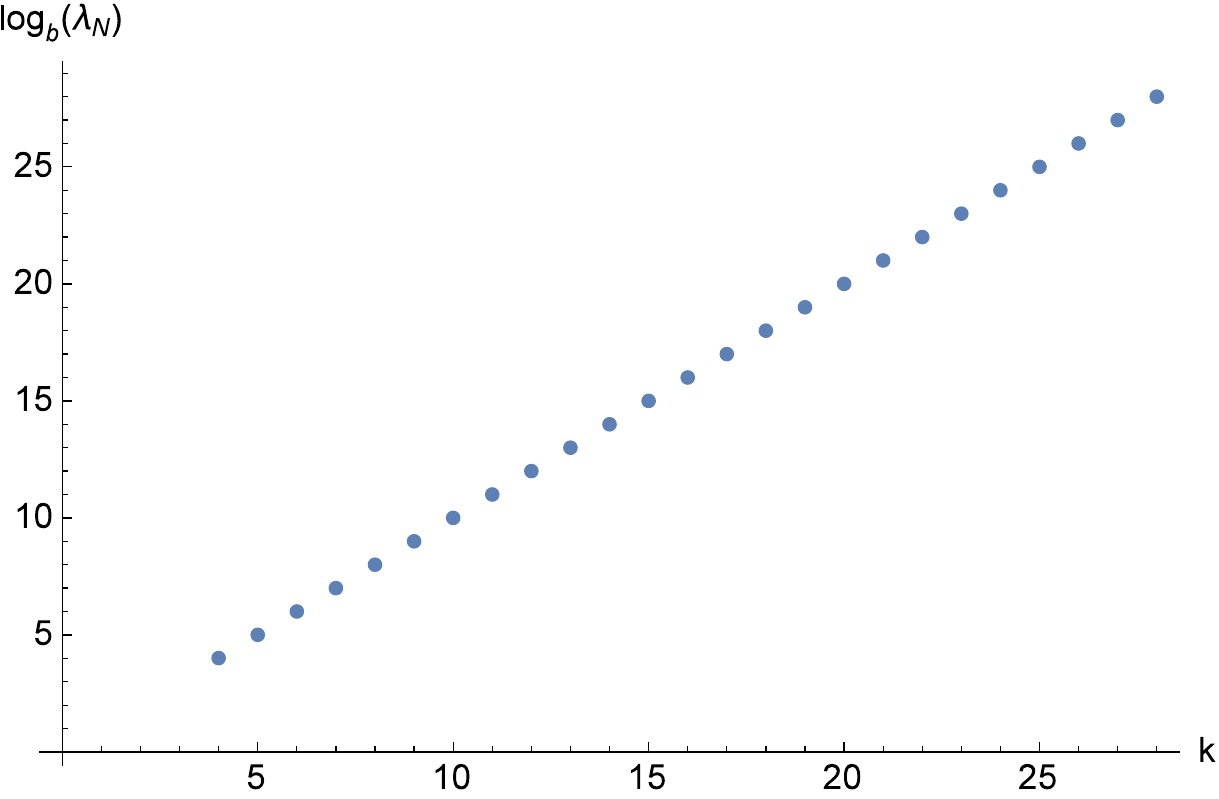}}

\subfloat[b=4]{\includegraphics[width=0.45\textwidth]{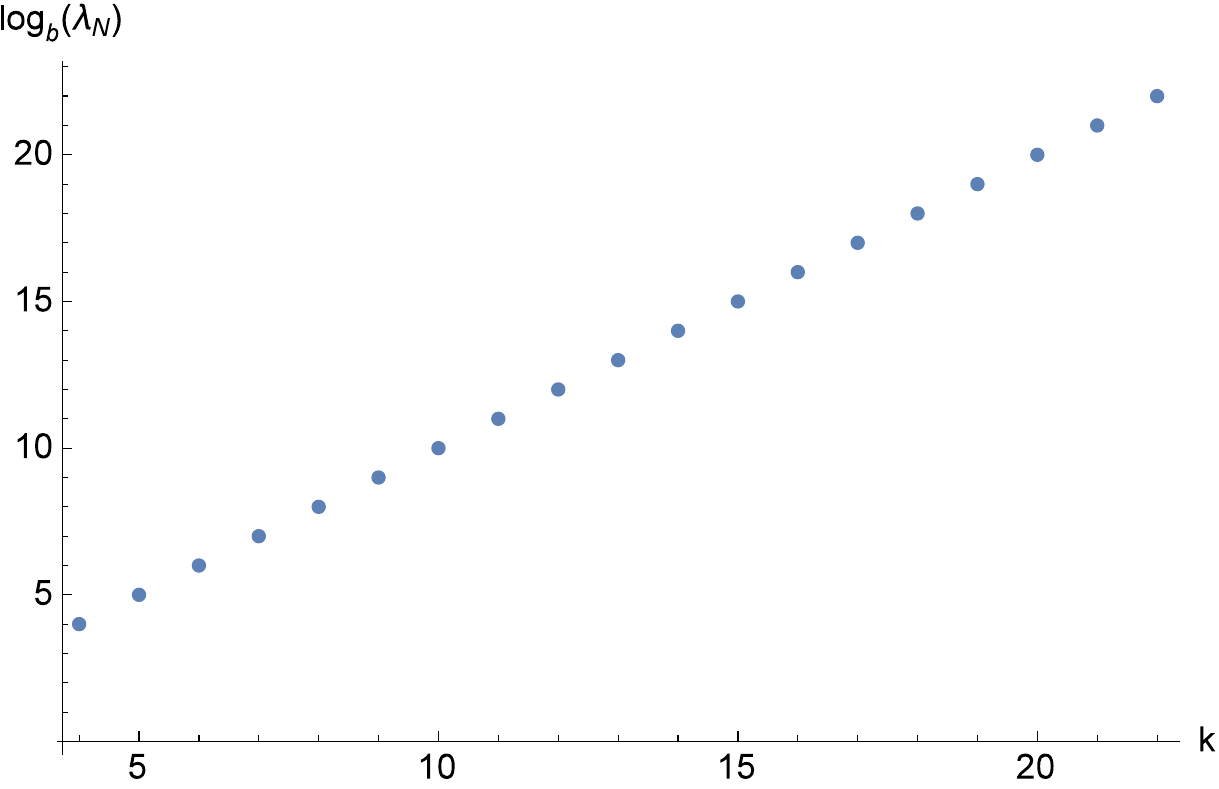}

}\qquad{}\subfloat[b=5]{\includegraphics[width=0.45\textwidth]{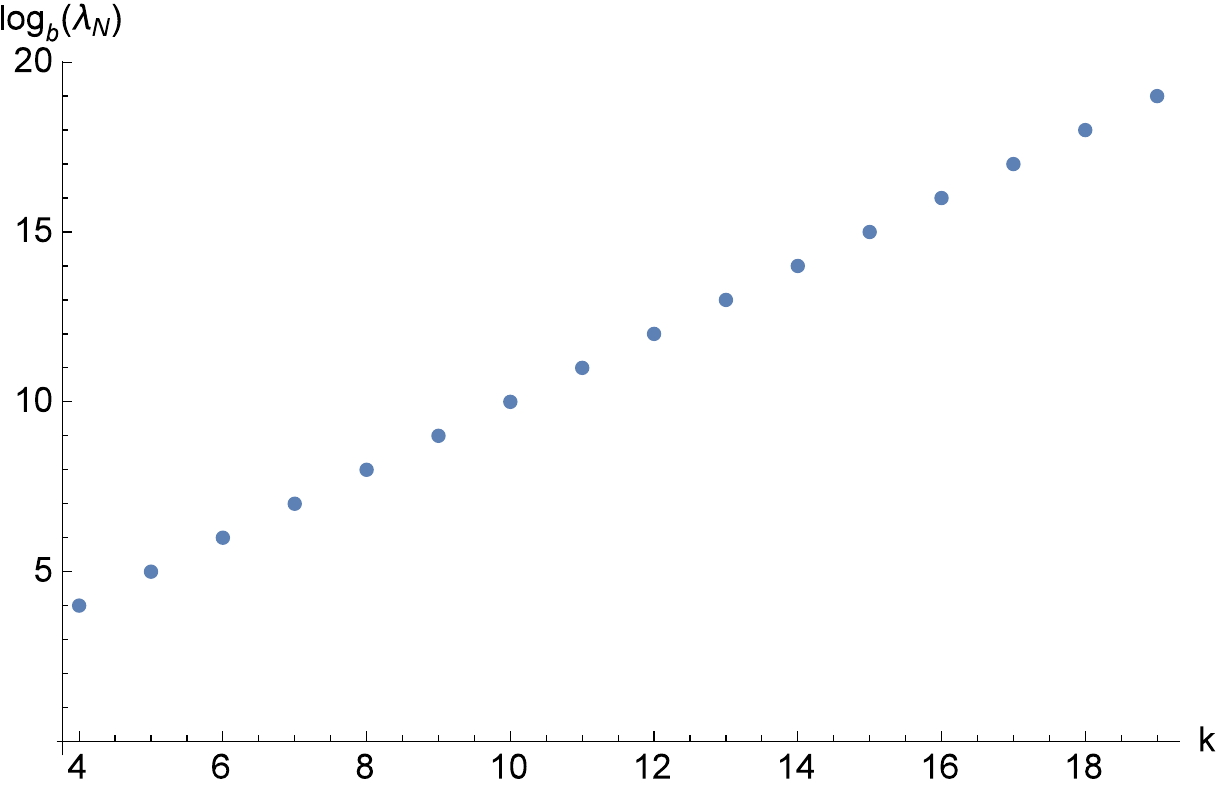}}

\caption{\label{fig:maxEigenvalue-MK}Plots of the largest eigenvalue $\lambda_{N}$
with system size $k(\sim\log_{b}N)$ for MKRG with $b=2,3,4,5$. It
can be fitted respectively as $\log_{b=2}\lambda_{N}=0.9992544987\ldots k+0.0213953207\ldots$,
$\log_{b=3}\lambda_{N}=0.9998501326\ldots k+0.0030754291\ldots$,
$\log_{b=4}\lambda_{N}=0.9999427813\ldots k+0.0009418968\ldots$,
and $\log_{b=5}\lambda_{N}=0.9999734834\ldots k+0.0003826594\ldots$,
showing that $\lambda_{N}\sim b^{k}\sim N$.}
\end{figure}

We have calculated the exact asymptotic scaling of spectral zeta-functions
for Hanoi networks and MKRG using the renormalization group. The results
highlight the importance of the spectral exponent $d_{s}$ for many
physical applications, such as synchronization and quantum searches.
For example, in Ref. \citep{LiBo16}, we use Eq.\ (\ref{eq:IjN})
to show that the efficiency of continuous-time quantum walks is controlled
by $d_{s}$ for any network, which generalizes the results previously
obtained for hyper-cubic lattices \citep{Childs04}. Synchronization
of identical dynamical systems in a network has been shown in Ref.
\citep{Barahona02} to depend on the scaling of the eigenratio of
largest to smallest nonzero eigenvalue of the network Laplacian. From
our analysis in Sec.\ \ref{sec:RG-Calculation} and Sec.\ \ref{sec:RG-for-MKRG},
the smallest nonzero eigenvalue $\lambda_{2}>0$ can be approximated
by $I_{1}$, when $d_{s}<2$ is satisfied. It is also suggested by
the argument in Sec.\ \ref{subsec:Heuristic-Arguments}. For HN3,
a degree-3 network, the largest eigenvalue is bounded above by $\lambda_{N}\leq6$,
and it is interesting to show (in the Appendix) how to use RG with
the ``power method'' \citep{kuczynski1992estimating} for matrices
to find that, in fact, $\lambda_{N}=5.37272879308215\ldots$. The
same arguments are applied to HN5, for which the asymptotic value
of $\lambda_{N}$ evolved with the numerical power method for varying
size $N=2^{k}$ is presented in Fig.~\ref{fig: maxEigenvalue-HN5}.
The largest eigenvalues are shown to scale with $\lambda_{N}\sim2\log_{2}N$.
This method also allows us to obtain the asymptotic value of $\lambda_{N}$
for hierarchical networks from MKRG, which scales as $\lambda_{N}\sim b^{k}\sim N$,
shown in Fig.~\ref{fig:maxEigenvalue-MK}. 

The calculation on the smallest nonzero and largest eigenvalue $\lambda_{2}$
and $\lambda_{N}$ allows us to analyze the synchronizability of all
the relevant networks. The linear stability of the synchronous state
is related to an algebraic condition of the Laplacian matrix according
to Ref. \citep{Barahona02}. The generic requirement for the synchronous
state to be linearly stable is $\sigma\lambda_{i}\in\left(\alpha_{1},\alpha_{2}\right)$
for all the nonzero eigenvalues of the Laplacian matrix, where $\sigma$
is the globle coupling, and $\left(\alpha_{1},\alpha_{2}\right)$
is the negative region of the master stability function that depends
solely on the dynamical system. For dynamical systems on network of
arbitrary topology, whether the network is synchronizable is decided
by the algebraic condition $\lambda_{N}/\lambda_{2}<\alpha_{2}/\alpha_{1}(=const)$.
This eigenratio determines the synchronizability of a network. The
eigenratios of HN3, HN5 and MKRG are asymptotically $N^{2-\log_{2}\phi}$,$2N\log_{2}N$
and $N^{\left(3+\log_{2}b\right)/\left(1+\log_{2}b\right)}$. This
would imply that the synchronizability for these networks is ranked
as HN5>HN3>MKRG, as long as $d_{s}<2/\left(1-\log_{2}\phi\right)\approx6.54$
(or $b<46.5$).

\paragraph*{Acknowledgements:}

We acknowledge financial support from the U. S. National Science Foundation
through grant DMR-1207431. 

\section*{Appendix: \label{sec:Appendix:}}

\subsection{Largest Eigenvalue\label{subsec:Largest-EV}}

We can use the power method, commonly used to numerically extract
particular eigenvalues of a matrix, to obtain the largest eigenvalue
of the Laplacian for HN3 analytically. The power method simply proceeds
as follows: Choose any generic vector ${\bf x}_{0}$ (that is non-zero
and not already an eigenvector associated with another eigenvalue),
then the evolution of
\begin{eqnarray}
{\bf x}_{t+1} & = & \frac{1}{\bar{\lambda}_{t}}{\bf L}{\bf x}_{t},\label{eq:powermethod}
\end{eqnarray}
converges to the eigenvector associated with the (absolute) largest
eigenvalue (if unique) of any matrix ${\bf M}$, where 
\begin{eqnarray}
\bar{\lambda_{t}} & = & \left\Vert {\bf x}_{t}\right\Vert \label{eq:norm}
\end{eqnarray}
ensures proper normalization of the evolving vector ${\bf x}_{t}$.
The magnitude of that largest eigenvalue is provided by $\lambda_{N}=\lim_{t\to\infty}\bar{\lambda}_{t}$.
Hence, analytically, we are faced with solving the fixed point equation
\begin{eqnarray}
{\bf x}^{*} & = & \frac{1}{\lambda_{N}}{\bf L}{\bf x}^{*},\label{eq:PMfixedpoint}
\end{eqnarray}
which is typically hopeless in general. But in case of the very sparse,
hierarchical Laplacian matrix for HN3, the set of $N=2^{k}$ coupled
linear equations defined by Eq.\ (\ref{eq:PMfixedpoint}) can be
solved again recursively. Then, we can write for Eq.\ (\ref{eq:PMfixedpoint}):
\begin{eqnarray}
0 & = & \left(3-\lambda_{N}\right)x_{0}-x_{2^{k}-1}-x_{1}-x_{2^{k-1}},\nonumber \\
0 & = & \left(3-\lambda_{N}\right)x_{2^{k-1}}-x_{2^{k-1}-1}-x_{2^{k-1}+1}-x_{0},\label{eq:PMfpHN3}\\
0 & = & \left(3-\lambda_{N}\right)x_{2^{i-1}(4j-3)}-x_{2^{i-1}(4j-3)-1}-x_{2^{i-1}(4j-3)+1}-x_{2^{i-1}(4j-1)},\nonumber \\
0 & = & \left(3-\lambda_{N}\right)x_{2^{i-1}(4j-1)}-x_{2^{i-1}(4j-1)-1}-x_{2^{i-1}(4j-1)+1}-x_{2^{i-1}(4j-3)},\nonumber 
\end{eqnarray}
for all $1\leq i<k$ and $1\leq j\leq2^{i-2}$. The recursion consists
of solving for and eliminating all odd-index ($i=1$) variables. To
that end, we re-write Eqs.\ (\ref{eq:PMfpHN3}) as
\begin{eqnarray}
0 & = & q_{2}\,x_{n-2}-p\left(x_{n-3}+x_{n-1}\right)-l\left(x_{n-4}+x_{n}\right)-x_{+},\nonumber \\
0 & = & q_{1}\,x_{n-1}-p\left(x_{n-2}+x_{n}\right)-x_{n+1},\label{eq:PMRGHN3}\\
0 & = & q_{2}\,x_{n}-p\left(x_{n-1}+x_{n+1}\right)-l\left(x_{n-2}+x_{n+2}\right)-x_{n\pm4},\nonumber \\
0 & = & q_{1}\,x_{n+1}-p\left(x_{n+2}+x_{n}\right)-x_{n-1},\nonumber \\
0 & = & q_{2}\,x_{n+2}-p\left(x_{n+3}+x_{n+1}\right)-l\left(x_{n+4}+x_{n}\right)-x_{-},\nonumber 
\end{eqnarray}
 for all $n=2(2j-1)$, $j=1,\ldots,2^{k-2}$, where initially
\begin{eqnarray}
q_{1}^{(0)}=q_{2}^{(0)} & = & 3-\lambda_{N},\nonumber \\
p^{(0)} & = & 1,\label{eq:RGinit}\\
l^{(0)} & = & 0.\nonumber 
\end{eqnarray}
Solving for and eliminating all odd-indexed variables $x_{n\pm1}$,
we find
\begin{eqnarray}
0 & = & \left(q_{2}-\frac{2p^{2}q_{1}}{q_{1}^{2}-1}\right)\,x_{n-2}-\left(l+\frac{p^{2}}{q_{1}-1}\right)\left(x_{n-3}+x_{n-1}\right)-\frac{p^{2}}{q_{1}^{2}-1}\left(x_{n-4}+x_{n}\right)-x_{+},\nonumber \\
0 & = & \left(q_{2}-\frac{2p^{2}}{q_{1}-1}\right)\,x_{n}-\left(l+\frac{p^{2}}{q_{1}-1}\right)\left(x_{n-2}+x_{n+2}\right)-x_{n\pm4},\label{eq:RGafter}\\
0 & = & \left(q_{2}-\frac{2p^{2}q_{1}}{q_{1}^{2}-1}\right)\,x_{n+2}-\left(l+\frac{p^{2}}{q_{1}-1}\right)\left(x_{n+3}+x_{n+1}\right)-\frac{p^{2}}{q_{1}^{2}-1}\left(x_{n+4}+x_{n}\right)-x_{-}.\nonumber 
\end{eqnarray}
Similar to the renormalization group treatment of HN3 in Sect. \ref{sec:RG-Calculation},
we relabel $x'_{n'+2}=x_{n+2}$, $x'_{n'+1}=x_{n}$, and $x'_{n'}=x_{n-2}$,
and obtain
\begin{eqnarray}
q'_{1} & = & q_{2}-\frac{2p^{2}}{q_{1}-1},\nonumber \\
q'_{2} & = & q_{2}-\frac{2p^{2}q_{1}}{q_{1}^{2}-1},\nonumber \\
p' & = & l+\frac{p^{2}}{q_{1}-1},\label{eq:PMrecur}\\
l' & = & \frac{p^{2}}{q_{1}^{2}-1},\nonumber 
\end{eqnarray}
considering which should be compared with Eqs.\ (\ref{eq:RG-HN3_redux}).
Then, Eqs.\ (\ref{eq:RGafter}) in terms of the primed quantities
take on exactly the form of the (lower three) Eqs.\ (\ref{eq:PMRGHN3})
and the circle closes. The recursion terminates after $k-2$ steps
with the equations
\begin{eqnarray}
0 & = & q_{2}^{(k-2)}\,x_{0}-p^{(k-2)}\left(x_{1}+x_{3}\right)-\left(2l^{(k-2)}+1\right)x_{2},\nonumber \\
0 & = & q_{1}^{(k-2)}\,x_{1}-p^{(k-2)}\left(x_{0}+x_{2}\right)-x_{3},\label{eq:PMfinal}\\
0 & = & q_{2}^{(k-2)}\,x_{2}-p^{(k-2)}\left(x_{1}+x_{3}\right)-\left(2l^{(k-2)}+1\right)x_{0},\nonumber \\
0 & = & q_{1}^{(k-2)}\,x_{3}-p^{(k-2)}\left(x_{0}+x_{2}\right)-x_{1},\nonumber 
\end{eqnarray}
which lead to the constraint
\begin{eqnarray}
0 & = & q_{2}^{(k-2)}-2l^{(k-2)}+1.\label{eq:PMconstraint}
\end{eqnarray}
Combining Eqs.\ (\ref{eq:RGinit}), (\ref{eq:PMrecur}), and (\ref{eq:PMconstraint})
provide an efficient procedure to determine the largest eigenvalue
$\lambda_{N}$, albeit implicit. For instance, for $k=2$, we can
directly insert Eqs.\ (\ref{eq:RGinit}) into Eq.\ (\ref{eq:PMconstraint})
to find $\lambda_{4}^ {}=4$, for $k=3$, we recur the initial conditions
in Eqs.\ (\ref{eq:RGinit}) once through Eqs.\ (\ref{eq:PMrecur})
before we apply the constraint in Eq.\ (\ref{eq:PMconstraint}) to
get 
\begin{eqnarray*}
0 & = & \left(3-\lambda_{8}\right)-\frac{2\left(3-\lambda_{8}\right)}{\left(3-\lambda_{8}\right)^{2}-1}+\frac{2}{\left(3-\lambda_{8}\right)^{2}-1}+1
\end{eqnarray*}
with the solution $\lambda_{8}=4+\sqrt{2}=5.414\ldots$. Beyond that,
a closed-form solution becomes quite difficult, and we have to resort
to an implicit ``shooting'' procedure, which is nonetheless exponentially
more efficient, $O(k=\log_{2}N)$, than a numerical evaluation with
the power method: simply choose a trial value for $\lambda_{N}$ in
Eqs.\ (\ref{eq:RGinit}) and evolve the recursion in Eqs.\ (\ref{eq:PMrecur})
until the right-hand side of Eq.\ (\ref{eq:PMconstraint}) has sufficiently
converged, then vary the value of $\lambda_{N}$ (using bisectioning
or regula-falsi) such that the constraint in Eq.\ (\ref{eq:PMconstraint})
is ever-better satisfied. In this way, we find 
\begin{eqnarray}
\lambda_{N} & = & 5.37272879308215\ldots,\label{eq:largestEV}
\end{eqnarray}
where in the end we need to evolve the recursions in Eq.\ (\ref{eq:PMrecur})
nearly 50 times before we can discern the convergence of the constraint.
This corresponds to an accuracy in the asymptotic value of $\lambda_{N}$
that would have required to evolve with the numerical power method
the Laplacian for HN3 of size $N=2^{50}$.

Similar to HN3, the renormalization group treatment with power method
is also applied to HN5. We obtain the recursions as 

\begin{eqnarray}
q' & = & q+2l-\frac{2p^{2}}{q-1},\nonumber \\
r' & = & r-\frac{2p^{2}q}{q^{2}-1},\nonumber \\
p' & = & l+\frac{p^{2}}{q-1},\label{eq:PMrecur-Hn5}\\
l' & = & 1+\frac{p^{2}}{q^{2}-1},\nonumber 
\end{eqnarray}
 the initial condition is 

\begin{eqnarray}
q^{\left(0\right)} & = & 3-\lambda_{N},\nonumber \\
r^{\left(0\right)} & = & 2\thinspace k-\lambda_{N},\\
p^{(0)} & = & 1,\label{eq:RGinit-HN5}\\
l^{(0)} & = & 1.\nonumber 
\end{eqnarray}
 The recursion terminates after $k-2$ steps with the equations
\begin{eqnarray}
0 & = & r^{\left(k-2\right)}\,x_{0}-p^{(k-2)}\left(x_{1}+x_{3}\right)-2l^{(k-2)}x_{2},\nonumber \\
0 & = & q^{\left(k-2\right)}\,x_{1}-p^{(k-2)}\left(x_{0}+x_{2}\right)-x_{3},\label{eq:PMfinal-HN5}\\
0 & = & r^{\left(k-2\right)}\,x_{2}-p^{(k-2)}\left(x_{1}+x_{3}\right)-2l^{(k-2)}x_{0},\nonumber \\
0 & = & q^{\left(k-2\right)}\,x_{3}-p^{(k-2)}\left(x_{0}+x_{2}\right)-x_{1},\nonumber 
\end{eqnarray}
which lead to the constraint
\begin{eqnarray}
0 & = & r^{\left(k-2\right)}+2l^{(k-2)}.\label{eq:PMconstraint-HN5}
\end{eqnarray}

Same method also apply to MKRG, in which the recursions 

\begin{eqnarray}
q_{i}' & = & q_{i+1}-2\frac{p^{2}}{q_{0}},\nonumber \\
p' & = & \frac{p^{2}}{q_{0}},\label{eq:PMrecur-MK}
\end{eqnarray}
 the initial condition is 

\begin{eqnarray}
q_{i}^{\left(0\right)} & = & 2-\lambda_{N}/b^{i},\qquad0\leq i<k,\nonumber \\
q_{k}^{\left(0\right)} & = & 2-2\lambda_{N}/b^{k},\\
p^{(0)} & = & 1.\label{eq:RGinit-HN5-1}
\end{eqnarray}
 The recursion terminates after $k$ steps with the equations
\begin{eqnarray}
0 & = & b^{k}\left[q^{\left(k\right)}/2\,x_{0}-p^{(k)}x_{1}\right],\nonumber \\
0 & = & b^{k}\left[q^{\left(k\right)}/2\,x_{1}-p^{(k)}x_{0}\right],\label{eq:PMfinal-MK}
\end{eqnarray}
which lead to the constraint
\begin{eqnarray}
0 & = & q^{\left(k\right)}/2+p^{(k)}.\label{eq:PMconstraint-MK}
\end{eqnarray}

\bibliographystyle{apsrev}
\bibliography{/Users/stb/Boettcher,SpectralZeta2-Ref-extra}

\end{document}